\documentclass[twocolumn,twoside,slac]{revtex4}
\usepackage{graphicx}
\usepackage{fancyhdr}
\begin{document}

\title{BOA: Framework for Automated Builds}

%

\author{N. Ratnikova}
\affiliation{FNAL, Batavia, IL 60510, USA}

\begin{abstract}
Managing large-scale software products is a complex software engineering task. The 
automation of the software development, release and distribution process is most 
beneficial in the large collaborations, where the big number of developers, multiple
 platforms and distributed environment are typical factors. 

This paper describes Build and Output Analyzer framework and its components that have
been developed in CMS to facilitate software maintenance and improve software quality. 

The system allows to generate, control and analyze various types of automated software 
builds and tests, such as regular rebuilds of the development code, software 
integration for releases and installation of the existing versions.
\end{abstract}

\maketitle

\thispagestyle{fancy}


\section{Introduction}
The Compact Muon Solenoid, CMS, detector on the Large Hadron Collider, LHC, is one 
of the largest international scientific collaborations. It unites almost 2 thousand 
scientists and engineers from 159 institutes in 36 countries of Europe, Asia, the 
Americas and Australasia.

CMS has a massive software development going on. Over 300 public software releases
were produced during last 15 months \cite{status}. This is a result of work of 
many developers world-wide.  CMS has a well-developed software infrastructure based 
on the Software Configuration, Release and Management tool SCRAM \cite{scram}, 
which has being adopted by the LHC Computing Grid project LCG at CERN at the end 
of last year.  

CMS SCRAM-managed Object Oriented projects currently include CMSToolBox, COBRA, 
FAMOS, Geometry, IGUANA, ORCA, and OSCAR. Projects release schedules and frequencies 
are flexible. Every project establishes release procedures, that are most convenient 
for the developers.  Due to cross-dependencies between projects and a big number of 
required external products and tools, considerable efforts are consumed to provide 
the consistency of the configuration requirements. The resulting system has a big 
number of configuration parameters, which may differ from site to site. Maintenance and 
support of the CMS software  installations and domain specific configuration
information becomes more and more challenging task. 

Goal of this project is to facilitate the software maintenance and help to improve 
software quality in the areas of software development, release management, software 
distribution and installation processes with the aids of automated builds.

The proposed Build and Output Analyzer framework BOA is intended to systematize
available tools and components of the existing CMS software infrastructure and
make them all to work together in a highly automated fashion. 

\section{Case Study}

CMS applies ongoing efforts in developing and customizing available automated tools 
and infrastructure for the software management. 

Whereas manual operations are still inevitable, most of repeatedly performed actions 
could be automated.  Software build
procedures are substantially automated with the aids of SCRAM native build system.
A centralized system of CVS repositories provides source code versions management 
and distribution. Software release events are automatically monitored by the 
ProjectWatch \cite{status} system. WarningFilter \cite{filter} tool allows to 
process software builds output and to publish statistics on the Web. 
Finally, software validation tool Oval \cite{oval} has been developed for the
detection of unexpected changes in the software behavior and control over its physics 
performances.

\begin{figure*}[t]
\centering
\includegraphics[width=135mm]{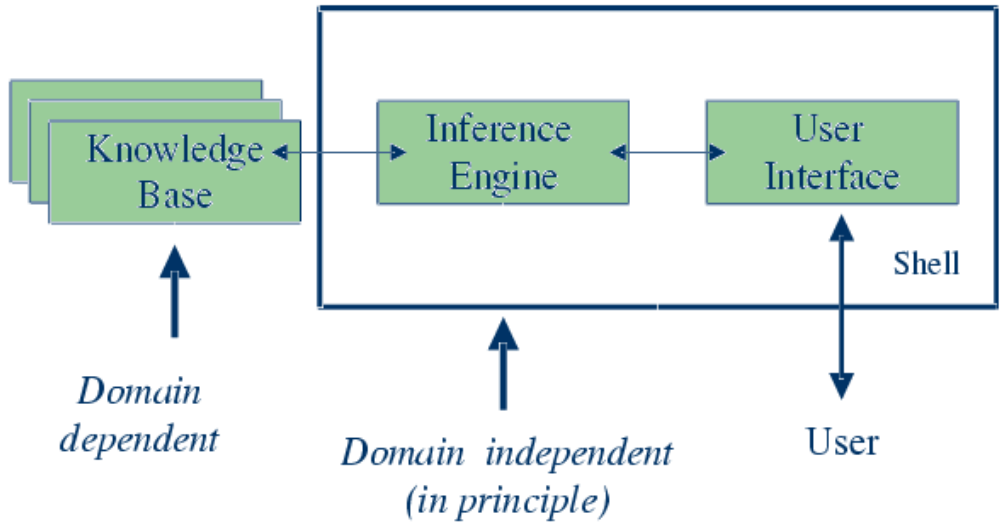}
\caption{BOA framework architecture corresponds to the Basic Expert System Architecture.} \label{Fig1}
\end{figure*}

Cross dependencies between projects are handled through the SCRAM configuration
mechanisms. Each project has a list of required tools, the number varying from a 
few to several dozens of tools. Tool configuration files are stored in a common 
repository. CMS  configuration releases provide consistent sets of tool versions for 
integration of the dependent projects. CMS software is supported for multiple 
computing platforms, including a range of versions of the operating system, and 
alternative compilers. Platform specific parameters contribute to overall 
configuration picture. 

Participating Regional Centers often do not have all required  external products and 
versions available. Users or local librarians need to install and maintain a big number of
supporting software packages and tools along with the proper CMS software projects. 
At the  same time they have to comply with the local administrative policies. Our 
experience shows, that site-specific configuration management involves most tedious, 
and error-prone manual operations. The software infrastructure and requirements are 
constantly evolving. The CMS baseline configuration tool SCRAM currently undergoes active 
development in order to provide improvements and extensions for the growing users base.

Note, that complex project dependencies, a big number of configurable parameters and 
frequent changes are intrinsic characteristics of the software process in any large HEP 
collaboration. 

\section{BOA Solutions}

The solutions discussed in this paper are mainly focused on the automation of the 
management of the CMS software domain configuration and workflow. This includes
distribution and installation of the external products and tools, configuration and 
build of the CMS software projects, analysis and validation tests of the build 
results. 

BOA model is based on the invariants, common to any complex software system. 
It uses the concepts of DOMAIN, PLATFORM, PROJECT, VERSION, INSTALLATION, 
and a range of related ACTIONS. 

This top-to-bottom approach allows to abstract the system structure from its 
functionality, and the functionality of the system from the particular implementation. 

BOA framework accumulates and systematizes the knowledge base for various operations, 
required for the software installation, successful software builds, and tests. 
In particular, the system keeps track of the current Domain configuration and the 
status of all builds. It provides standard interfaces to the underlying components 
and tools. In addition it allows to reuse common utilities, and the information 
available in the Domain.

\begin{figure*}[t]
\centering
\includegraphics[width=135mm]{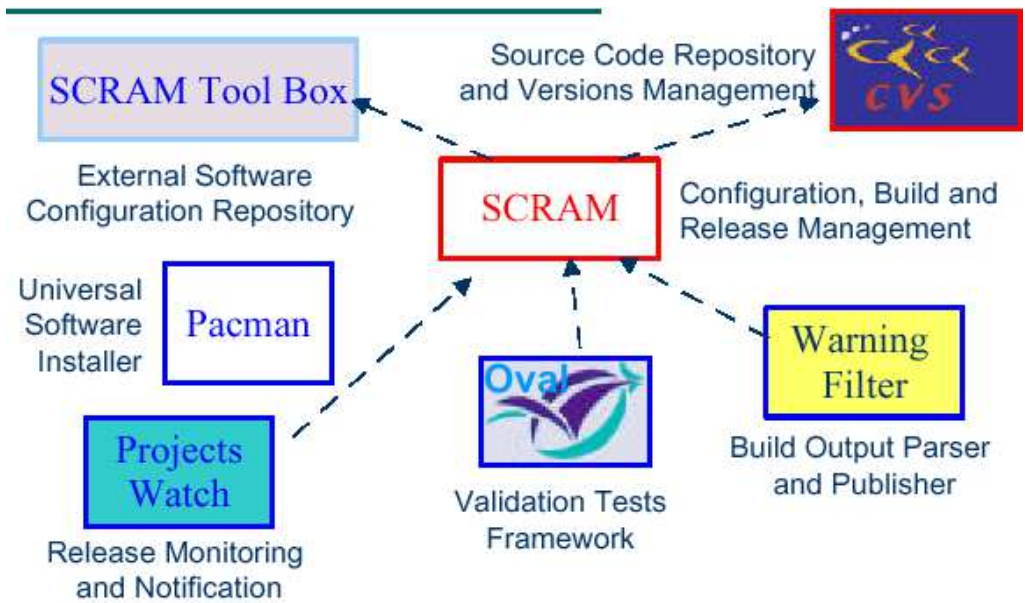}
\caption{Versatile automated services and tools are already available in CMS.} \label{fig1}
\end{figure*}

\section{Architecture and Components}

BOA design exploits the Object Oriented approach, main conceptual components being 
presented by following classes:
 
\begin{itemize}
\item DOMAIN is on the top of BOA structure. It contains site dependent information
and keeps a list of software Projects and Platforms. Domain takes care of installation 
and availability of the required tools, such as SCRAM, Pacman\cite{pacman}.

\item PROJECT carries information about a particular software product.
BOA currently supports two types of projects: ``scramified'' and ``pacmanized''.
The former specifies SCRAM managed projects installed via scram native 
bootstrap mechanisms and then built from sources. The latter specifies products 
distributed via Pacman caches. 
Pacman provides a convenient way for installation of the required external tools.
Project contains list of available versions, and other project specific information,
such as CVS repository or Pacman cache.

\item VERSION is responsible for the configuration requirements and installation specific 
information and algorithms. Versions keep track of the current status of the installations.

\end{itemize}

Entire domain information is stored persistently in the database. 
Framework can manipulate several domains at the same time. 
It can also work with different database instances.

\section{Features and Implementation Status}

One can work interactively in BOA environment, or run a predefined scenario without 
manual intervention. 

For the execution part BOA provides the Session class, which allows flexible control 
of the workflow. This is different from the widely used approach of generating scripts 
for the subsequent execution. The instance of the Session class opens a new OS shell 
process, where commands, defined in the framework,  are consecutively executed.  

The output and exit status of the commands are intercepted and can be analyzed. 
Depending on the results, program can choose the next step and execute it in the 
same environment. The requirement of unbroken environment is essential for most of 
software builds and test procedures, for instance the runtime environment needs to
be set prior the execution. Session keeps log of all actions. 

It is appropriate mention here, that the above approach is not applicable to the 
commands, that expect direct user's input. This special case should be handled 
differently. In general tools used in autonomous mode should be able to accept 
input as arguments. 

BOA interface is written it Python, and it is constantly evolving. Development includes a series 
of functioning prototypes, providing desired features and functionality bit by bit. 
Every micro-release passes a test suite. Unit tests are provided as well. 
New commands are  documented through the built-in help feature.

\section{Conclusions}

Described approach allows to address efficiently a wide complex of tasks of the
software librarian. At the same time proposed system does not dictate any specific
implementation, and successfully adopts tools chosen by the Collaboration.

BOA framework implementation offers a convenient solution for individuals responsible 
for configuring, upgrading, and supporting CMS software domain. 
\begin{widetext}

\end{widetext}  
\end{document}